\documentclass[letter]{aa}
\usepackage{graphicx}
\usepackage{txfonts}
\usepackage{aa}
\def\sn{\hbox{S/N}}  
  
\def\vsin{\hbox{$v \sin i$}}  
  
\def\kms{\hbox{km\,s$^{-1}$}}

\def\em{\it}

\begin{document}
   \title{Detection of a weak surface magnetic field on Sirius A: are all tepid stars magnetic ?\thanks{Based on observations obtained at the Bernard Lyot Telescope (TBL, Pic du Midi, France) of the Midi-Pyr\'en\'ees Observatory, which is operated by the Institut National des Sciences de l'Univers of the Centre National de la Recherche Scientifique of France, and at the Canada-France-Hawaii Telescope (CFHT) which is operated by the National Research Council of Canada, the Institut National des Sciences de l'Univers of the Centre National de la Recherche Scientifique of France, and the University of Hawaii.}}
   \titlerunning{The weak magnetic field of Sirius A }

   \author{
          P. Petit
          \inst{1,2}  
          \and
          F. Ligni\`eres
          \inst{1,2}
          \and
          M. Auri\`ere
          \inst{1,2}
          \and
          G.A. Wade
          \inst{3}
          \and
          D. Alina
          \inst{1,2}
          \and
          J. Ballot
          \inst{1,2}
          \and
          T. B\"ohm
          \inst{1,2}
          \and
          L. Jouve
          \inst{1,2}
          \and
          A. Oza
          \inst{1,2}
          \and
          F. Paletou
          \inst{1,2}
          \and
          S. Th\'eado
          \inst{1,2}
                    }

   \offprints{P. Petit}

   \institute{
   Universit\'e de Toulouse, UPS-OMP, Institut de Recherche en Astrophysique et Plan\'etologie, Toulouse, France \\ 
   \email{petit@ast.obs-mip.fr}
   \and
   CNRS, Institut de Recherche en Astrophysique et Plan\'etologie, 14 Avenue Edouard Belin, F-31400 Toulouse, France
   \and
   Department of Physics, Royal Military College of Canada, PO Box 17000, Station Forces, Kingston, Ontario, Canada \\ 
             }

   \date{Received ??; accepted ??}

 
  \abstract
   {}
   {We aim at a highly sensitive search for weak magnetic fields in main sequence stars of intermediate mass, by scanning classes of stars with no previously reported magnetic members. After detecting a weak magnetic field on the normal, rapidly rotating A-type star Vega, we concentrate here on the bright star Sirius A, taken as a prototypical, chemically peculiar, moderately rotating Am star.}
   {We employed the NARVAL and ESPaDOnS high-resolution spectropolarimeters to collect 442 circularly polarized spectra, complemented by 60 linearly polarized spectra. Using a list of about 1,100 photospheric spectral lines, we computed a cross correlation line profile from every spectrum, leading to a signal-to-noise ratio of up to 30,000 in the polarized profile.}
   {We report the repeated detection of circularly polarized, highly asymmetric signatures in the line profiles, interpreted as Zeeman signatures of a large-scale photospheric magnetic field, with a line-of-sight component equal to $0.2 \pm 0.1$ G.}
   {This is the first polarimetric detection of a surface magnetic field on an Am star. Using rough estimates of the physical properties of the upper layers of Sirius A, we suggest that a dynamo operating in the shallow convective envelope cannot account for the field strength reported here. Together with the magnetic field of Vega, this result confirms that a new class of magnetic objects exists among non Ap/Bp intermediate-mass stars, and it may indicate that a significant fraction of tepid stars are magnetic.}


   \maketitle

\section{Introduction}

If most high-resolution spectropolarimetric surveys of main-sequence A and late-B stars (sometimes called ``tepid'' stars) have been unsuccessful at detecting organized magnetic fields with longitudinal field strengths greater than a few gauss in objects that do not belong to the class of Ap/Bp stars (e.g. Shorlin et al. 2002, Wade et al. 2006, Auri\`ere et al. 2010, Makaganiuk et al. 2011), a sensitive spectropolarimetric investigation of the normal, rapidly-rotating A-type star Vega has recently revealed a weak magnetic field at its surface (Ligni\`eres et al. 2009). This first detection raises the question of the ubiquity of magnetic fields in objects belonging to this mass domain.

Sirius A is the brightest member of a visual binary system, the secondary component of which is a hot DA white dwarf. With an effective temperature of $9900\pm200$K (Lemke 1989), Sirius A is a hot Am (metallic-line) star, with non-solar abundances of various chemical species (Landstreet 2011).  Its diameter was estimated from near-infrared interferometry, with $D_A = 1.711 \pm 0.013 D_\odot$ (Kervella et al. 2003). Precise astrometry has allowed for accurately determining the masses of both components, with $M_A = 2.12 \pm 0.06 M_\odot$ and $M_B = 1.000 \pm 0.016 M_\odot$ (Gatewood \& Gatewood 1978, Kervella et al. 2003). The projected rotational velocity \vsin\  of Sirius A is equal to $16 \pm 1$ \kms\ (Royer et al. 2002). While its equatorial velocity $v_{\rm eq}$ is unknown, it should be less than 120 \kms, as slow rotation is a common property of Am stars (Abt 2009). This small to moderate rotational velocity differentiates Sirius A from Vega ($v_{\rm eq} > 170$ \kms, according to Takeda et al. 2008). Sirius is therefore well-suited to testing whether Vega-like magnetism is limited to rapidly rotating stars or if it is instead a more widespread phenomenon.

In the present article, we report on the outcome of a deep spectropolarimetric study of Sirius A and report evidence for detection of a weak surface magnetic field. In Section \ref{sec:data}, we describe the instrumental setup and the data set collected for this study. The detection of polarized signatures is presented in Section \ref{sec:zeeman} and attributed to the Zeeman effect. The characteristics of the large-scale magnetic field of Sirius are discussed in Section \ref{sec:discussion}, and first steps toward determining its physical origin are proposed.

\section{Observations and data analysis}
\label{sec:data}

\begin{figure}
\centering
\includegraphics[width=9cm]{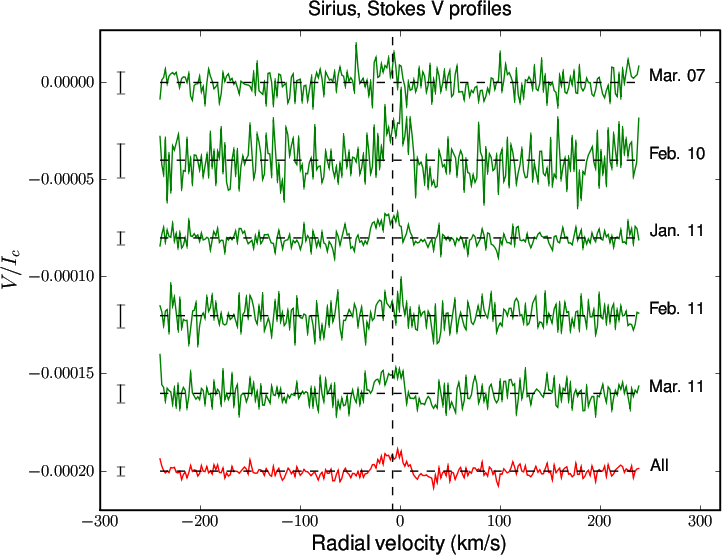}
\caption{{\bf Averaged} Stokes V LSD profiles of Sirius A for the successive epochs of observation (green lines). The red line is obtained by averaging all 442 Stokes V profiles. The profiles are successively shifted vertically for better clarity, with dashed lines indicating the zero level. The error bar corresponding to each profile is plotted on the left side of the plot. Vertical dashes show the stellar radial velocity.}
\label{fig:stokesv}
\end{figure}

\begin{figure}
\centering
\includegraphics[width=9cm]{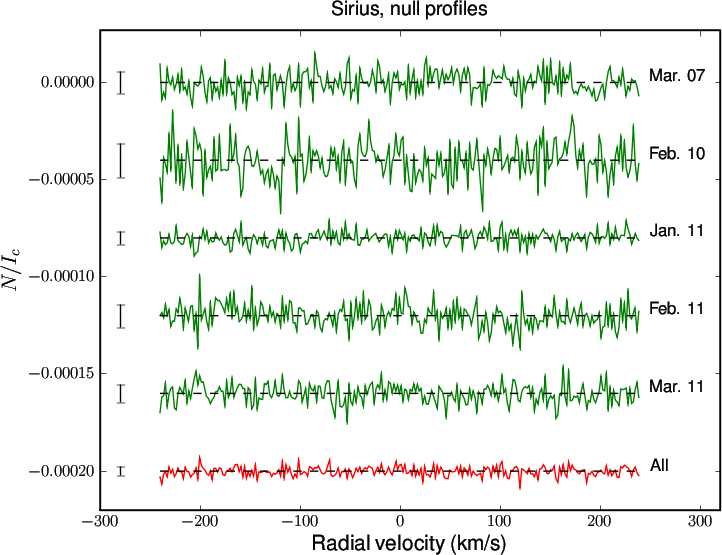}
\caption{``Null'' profiles obtained for the set of circularly polarized observations.}
\label{fig:null}
\end{figure}

\subsection{Instrumental setup, data reduction, and multi-line extraction of Zeeman signatures}

We describe here a series of spectropolarimetric observations of Sirius A obtained between 2007 and 2011 that used the NARVAL and ESPaDOnS twin stellar spectropolarimeters installed at T\'elescope Bernard Lyot (Pic du Midi Observatory, France) and Canada-France-Hawaii Telescope (Mauna Kea Observatory, Hawaii), respectively. The instrumental setup and reduction pipeline is identical to the one described by Petit et al. (2008), except for an upgrade of the CCD chip of ESPaDOnS in early 2011, enabling a gain in efficiency by a factor of about 2 in the red part of the spectrum, using the new detector (N. Manset, private communication). 

\begin{table}
\center
\caption[]{Journal of observations.}
\begin{tabular}{llllll}
\hline
Instrument & Date & Stokes   & no. & t$_{\rm exp}$ & \sn \\
                    &           &  param.  &  spectra & (sec)   &   (LSD) \\
\hline
NARVAL      & 12 Mar. 07      & V & 32  & $4\times 8.0$ & $31320 \pm 2407$ \\
ESPaDOnS & 02 Feb. 10      & V & 86  & $4\times 1.0$ & $11396 \pm 1527$ \\
NARVAL      & 24 Jan. 11 & V & 55  & $4\times 5.0$ & $30610 \pm 1405$ \\
NARVAL      & 25 Jan. 11 & V & 60  & $4\times 5.0$ & $27088 \pm 1753$ \\
NARVAL      & 04 Feb. 11 & Q & 30  & $4\times 5.0$ & $31381 \pm 1294$ \\
NARVAL      & 05 Feb. 11 & U & 30  & $4\times 5.0$ & $26465 \pm 2560$\\
ESPaDOnS & 19 Feb. 11 & V & 43 & $4\times 0.8$ & $20590 \pm 1506$ \\
ESPaDOnS & 20 Feb. 11 & V & 1 & $4\times 0.8$ & $14268$ \\
ESPaDOnS & 21 Feb. 11 & V & 37 & $4\times 0.8$& $18314 \pm 3967$ \\
ESPaDOnS & 14 Mar. 11 & V & 42 & $4\times 0.8$& $22943 \pm 943$ \\
ESPaDOnS & 15 Mar. 11 & V & 43 & $4\times 0.8$& $20520 \pm 1020$ \\
ESPaDOnS & 16 Mar. 11 & V & 43 & $4\times 0.8$& $16594 \pm 3921$ \\
\hline
\end{tabular}
\label{tab:obs}
\\
\end{table}

Each polarized spectrum is obtained from a combination of four subexposures taken with the half-wave rhombs oriented at different azimuths (Semel et al. 1993). The data reduction is automatically performed by a pipeline implementing the optimal spectral extraction principle of Horne (1986) and Marsh (1989). All spectra are processed using the least-squares-deconvolution technique (LSD hereafter, Donati et al. 1997), to extract a mean line profile with enhanced \sn. The line list employed here is computed from a Kurucz atmospheric model with a surface effective temperature T$_{\rm eff} = 10,000$~K, surface gravity $\log(g)=4.0$, and solar metallicity. Using all 1,100 spectral lines listed in the model, the noise level of the LSD polarized profiles (listed in Table \ref{tab:obs}) is typically decreased by a factor of 10, compared to the peak \sn\ of the full spectrum. 

For the present analysis, we have collected a total of 502 spectra, with 442 circularly polarized (Stokes V) spectra and 60 linearly-polarized spectra equally split between the Stokes Q and U parameters. The series of 32 observations collected in March 2007 have already been presented by Auri\`ere et al. (2010). From our original data set we have only kept observations with an \sn\ of the polarized LSD profiles greater than 10,000. This selection criterion ensures that we exclude all data obtained at a high airmass, observed through thin clouds, or suffering from occasional inaccuracies in the telescope tracking.

\subsection{Zeeman signatures}
\label{sec:zeeman}

In spite of using a cross-correlation procedure, individual Stokes V LSD profiles do not exhibit any detectable polarized signature (circular or linear). To increase the \sn\ further, we calculated a series of temporal averages of LSD line profiles collected during close nights, as illustrated in Figure \ref{fig:stokesv}. We have to bear in mind that, by doing so, we add up observations taken at different rotation phases of the star and therefore extract only a phase-averaged signal. This strategy is, however, successful at reducing the noise level sufficiently (down to $2\times 10^{-6}I_c$, where $I_c$ is continuum level, when all observations are averaged together) to permit the detection of circularly polarized signatures at several epochs at the radial velocity of the star, equal to -7.4 \kms. Grouping all Stokes V data together enables us to reach a false alarm probability of the detected signal of $6\times 10^{-5}$, with a signal amplitude of $2\times 10^{-5}I_c$. Epoch-to-epoch comparisons of the signatures do not reveal any statistical differences. The signatures are strongly asymmetric about the line centre (while Stokes I asymmetry is less pronounced, see Fig. 1 of Auri\`ere et al. 2010), with a blue positive lobe dominating the signal, while a negative lobe is marginally detected in the red wing of the line profiles. The centre-of-gravity technique (Rees \& Semel 1979) provides us with an estimate of the longitudinal field strength from Stokes I and V profiles, with $B_{\rm eff} = 0.2 \pm 0.1$G (using all data together). We stress, however, that this method is based on assuming a  Stokes V signature of zero integral, so that its outcome in the case of a strongly distorted Stokes V profile shape (as illustrated by the prominent blue lobe observed here) should be taken with caution and will most likely underestimate the line-of-sight field strength (a bias reinforced by our estimate being phase-averaged and therefore limited to the global, axi-symmetric field component).

\begin{figure}
\centering
\includegraphics[width=9cm]{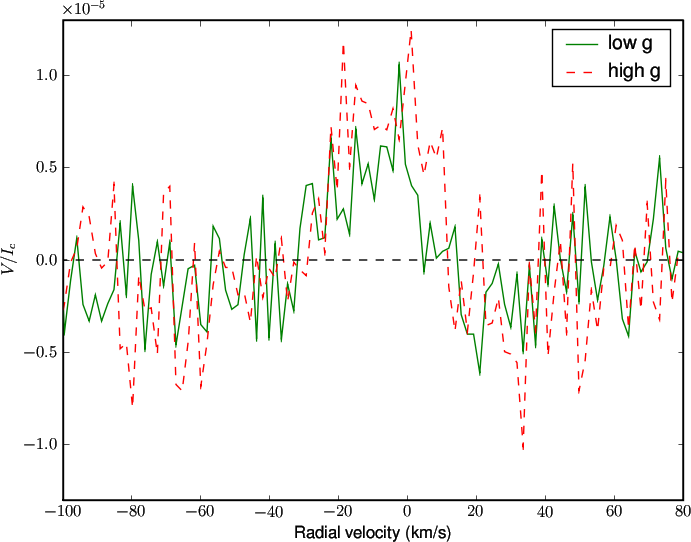}
\caption{Comparison of the Stokes V profiles obtained by selecting photospheric lines of low (green line) and high magnetic sensitivity (red curve).}
\label{fig:lande}
\end{figure}

\begin{figure}
\centering
\includegraphics[width=9cm]{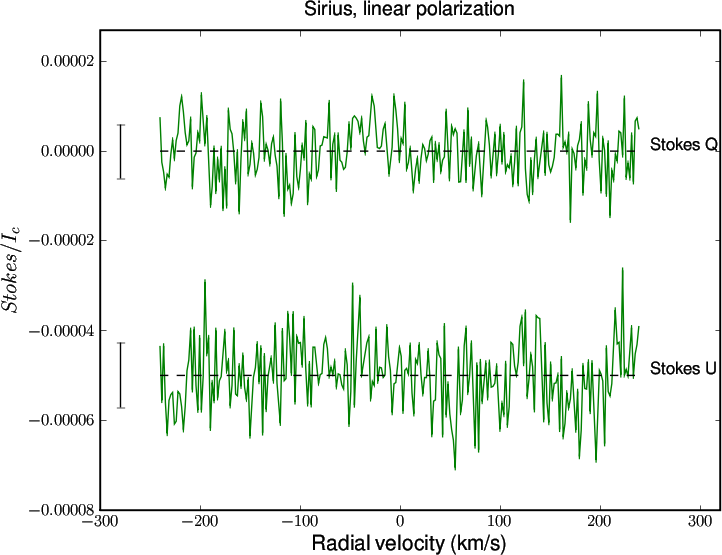}
\caption{Averaged Stokes Q and U LSD profiles.}
\label{fig:stokesqu}
\end{figure}

By computing a different combination of the 4 subexposures collected to produce a polarized spectrum, it is possible to calculate a ``null'' diagnostic spectrum  (Semel et al. 1993), which should not contain any polarized stellar signal and is indicative, in the case of a non-zero profile, of possible artefacts in the measurements of Stokes parameters of instrumental and/or stellar origin (e.g. stellar pulsations). The null profiles are represented in Figure \ref{fig:null}. Their flat shapes provide us with an additional argument favouring a signal of stellar origin. To further evaluate the proposed Zeeman nature of the signatures, we run again the cross-correlation procedure for two new line lists, defined in our original list and containing lines with an average Land\'e factor $g$ lower (resp. greater) than 1.2, with mean $g$ values of 0.94 and 1.51. Since spectral lines with higher $g$ values are more sensitive to the Zeeman effect, we expect a higher amplitude of their polarized signatures in case of a signal of magnetic origin (assuming a similar mean wavelength for the line lists with low and high $g$, which is actually the case here). The outcome of this test is illustrated in Figure \ref{fig:lande}, after correcting the Stokes V profiles for a $\approx 10$\% difference in equivalent width between their associated Stokes I profiles. The high-$g$ profile has a marginally higher amplitude than its low-$g$ counterpart, with an amplitude ratio roughly consistent with the ratio of the $g$ values. However, this difference cannot be taken as statistically significant, given the relative noise. As another test, we obtained a set of Stokes Q and U observations, with the corresponding averaged LSD profiles plotted in Figure \ref{fig:stokesqu}. With an \sn\ of $\approx 10^{-5}I_c$ in linear polarization, the absence of any signal detection is consistent with a magnetic origin of the signature in Stokes V, since Stokes Q/U Zeeman signatures are much weaker than their Stokes V counterpart (Wade et al. 2000). The absence of linear signatures also indicates that the Stokes V signature is not the result of any crosstalk from linear to circular polarization. 

We finally estimated the possibility of a contamination of the polarized signature detected on Sirius A by its compact companion. Although Sirius B is not known to be a magnetic object, a small fraction of white dwarfs do host magnetic fields of up to $10^9$G (Liebert et al. 2005). Sirius B is classified as a DA white dwarf, implying that its spectral lines are almost strictly limited to hydrogen lines (Barstow et al. 2005). Hydrogen transitions are not included in the line-list employed here to compute the LSD profiles of Sirius A, so that a spurious detection caused by these specific spectral features is unlikely (in either case, the number of hydrogen lines recorded in NARVAL spectra of Sirius A is negligible compared to the number of metallic lines of our line mask). Another type of contamination could come from the magnetic field of Sirius B reaching the surface of its more massive companion and generating a detectable Zeeman effect in Sirius A. Given the rapid decrease in magnetic field strength above the surface of a magnetic star (proportional to $d^{-3}$ in the case of a dipole, where $d$ is the distance to the magnetic star), even an extremely strong field at the surface of Sirius B would not influence the line formation in the atmosphere of Sirius A in any detectable way, given the separation ranging between 8 and 32 AU between the two components of the binary system along their eccentric orbits (van den Bos 1960).   

Considering all above arguments together, we conclude that the observed cirularly polarized signature is most very likely due to the existence of a weak surface magnetic field on Sirius A.

\section{Discussion}
\label{sec:discussion}

The detection of a sub-gauss magnetic field on Sirius A is the second such detection on a tepid star (after Vega, Ligni\`eres et al. 2009), and the first one on an Am star. The shapes of Stokes V profiles recorded for both stars clearly differ and are indicative of different photospheric field properties. Stokes V signatures detected for Vega appear in the very core of the line profile, which was interpreted by Ligni\`eres et al. (2009) as a polar concentration of the magnetic flux, later confirmed through the modelling of the field topology by means of the Zeeman-Doppler imaging technique (Petit et al. 2010). The signature recovered for Sirius is broader, spanning most of the line width, which suggests a global distribution of the large-scale field. The rotation period of Sirius A is unknown, but its Am classification points towards a period longer than one day, i.e. likely longer than that of Vega (Takeda et al. 2008, Petit et al. 2010). The detection of a magnetic field on Sirius A therefore suggests that weak surface fields are not the prerogative of rapidly rotating tepid stars and may concern a significant fraction of objects in this mass domain. 

If the approximate surface field strength of 0.2~G derived in Sect. \ref{sec:zeeman} is far below equipartition (with an equipartition limit of the order of 100~G for the surface temperature of Sirius, Makaganiuk et al. 2011), this phase-averaged estimate carries selective information about the largest-scale magnetic field component and does not rule out the possibility of a tangled surface field, which could locally reach a higher strength (within the tight upper limit of 2~G derived from individual spectra). 

The predominance of the blue lobe in the Stokes V signatures of Sirius A, resulting in a non-zero integral of Stokes V over the line width, is not expected in the standard theory of the Zeeman effect. This abnormal profile shape may be an extreme example of a phenomenon that is not only well-documented by solar physicists (e.g. Viticchi\'e \& Sanchez Almeida 2011 and references therein), but also observed in cool stars (Petit et al. 2005, Auri\`ere et al. 2008). In the solar context, Stokes V asymmetry is generally interpreted as a combination of vertical gradients in both velocity and magnetic fields inside magnetic elements (e.g. Lopez Ariste 2002 and references therein). The convective motions in the upper layers of Sirius A can be inferred from detection of a micro-turbulence contribution to the broadening of spectral lines (Landstreet et al. 2009). Strongly distorted Stokes V profiles may therefore be linked to the velocity and magnetic gradients related to the convective mixing. Recent hydrodynamical models predicting the widespread presence of shocks in these superficial layers (Kupka et al. 2009) give some support to this preliminary interpretation. Given the big differences in physical properties between Sirius A and the Sun, we stress, however, that a separate investigation of this phenomenon is obviously required before reaching any conclusion.  

Observing this possible imprint of convection on the magnetic signal has led us to investigate whether the photospheric field can be generated within the thin convective envelope of Sirius A, through a solar-type dynamo. To do so, we used an approach similar to the one of Christensen et al. (2009), who derive a simple scaling law between the local dynamo output and the available convective energy flux of a spherical shell. Using estimates of the temperature scale height and density of the convective shell provided by a stellar evolutionary model of Sirius A (including atomic diffusion) computed with the TGEC code (Richard et al. 2004, Hui Bon Hoa 2008) and combining these values with the convective energy flux taken from Kupka et al. (2009), we derived an upper limit of about 0.1~G on a dynamo-generated local field. Although this value might seem, at first glance, roughly consistent with the value observationally derived for Sirius A, we have to recall that our estimate of the field strength has very likely been underestimated (see Sect. \ref{sec:zeeman}). We also stress that the method proposed by Christensen is only valid for a saturated dynamo and will overestimate the field in the case of a star in the non-saturated regime, which is likely the case for Sirius A. In this context, we tentatively conclude that the observed magnetic field cannot be produced in the shallow convective envelope. It may rather be stored in the stable layers of the star and thread the thin subsurface convective zone, up to the photosphere. This conclusion is supported by the absence of any detectable variability in the polarized signature over the four-yer timespan of our data set, while faster variability would be expected with a classical dynamo. Future observational and modelling work is, however, necessary to confirm this preliminary interpretation and help distinguish between the alternate options of either a field of fossil origin (Moss 2001), the weak magnetic remnant of a stronger field destroyed by the Tayler instability (Auri\`ere et al. 2007), or a field generated through a dynamo active either in the convective core (Brun et al. 2005, Featherstone et al. 2009) or in the radiative layers (Spruit 2002, Zahn et al. 2007).

\begin{acknowledgements}
We are grateful to the staffs of TBL and CFHT for their very efficient help during data collection. GAW acknowledges Discovery Grant support from the Natural Sciences and Engineering Research Council of Canada.\end{acknowledgements}

\end{document}